\documentclass[10pt,twocolumn,reprint,amsmath,amssymb,aps,pra,showpacs,longbibliography,superscriptaddress]{revtex4-1}
\usepackage[utf8]{inputenc}
\usepackage{amsmath}
\usepackage{amssymb}
\usepackage{graphicx}
\usepackage{xcolor}

\makeatletter
\usepackage{amsfonts}
\usepackage{graphicx}
\usepackage{graphics}

\makeatother

\begin{document}
\title{Properties of Bose-Einstein condensates with altermagnetism}
\author{Jia Wang}
\affiliation{Centre for Quantum Technology Theory, Swinburne University of Technology,
Melbourne 3122, Australia}
\author{Zhao Liu}
\affiliation{Centre for Quantum Technology Theory, Swinburne University of Technology,
Melbourne 3122, Australia}
\author{Xia-Ji Liu}
\affiliation{Centre for Quantum Technology Theory, Swinburne University of Technology,
Melbourne 3122, Australia}
\author{Hui Hu}
\affiliation{Centre for Quantum Technology Theory, Swinburne University of Technology,
Melbourne 3122, Australia}
\date{\today}
\begin{abstract}
  We investigate a weakly interacting two-component Bose--Einstein condensate in the miscible regime in the presence of \emph{altermagnetism}, i.e., a collinear and globally compensated magnetic order that breaks spin-rotation symmetry while maintaining zero net magnetization. Within Bogoliubov theory, we derive the quasiparticle spectrum and coherence factors and show that altermagnetic order generically induces an angular dependence of the low-energy excitations. As a result, the sound velocity, momentum-resolved magnetization in the quantum depletion, and density--spin response functions acquire anisotropic components. We show that these anisotropic contributions vanish after angular averaging, consistent with the defining feature of altermagnetism: nontrivial local spin polarization without a global magnetization. Finally, we evaluate the Lee--Huang--Yang correction to the ground-state energy in the altermagnetic phase. Our results should be testable with ultracold-atom experiments in the foreseeable future.
\end{abstract}
\maketitle

\section{Introduction}

Ultracold atoms provide a versatile platform for quantum simulation, offering unprecedented control over dimensionality, interactions, geometry, and disorder \cite{Bloch2008RMP,Lewenstein2007AdvPhys,Georgescu2014RMP}. This high degree of tunability enables clean and quantitative realizations of paradigmatic many-body Hamiltonians and has provided insights into central condensed-matter mechanisms, including superconductivity and Bose-Einstein condensation, magnetism, charge orders, and the associated collective excitations \cite{Bloch2008RMP,Georgescu2014RMP}. Motivated by this broader program, here we explore how a recently introduced concept from solid-state physics---\emph{altermagnetism}---can be implemented and diagnosed in a cold-atom setting. Altermagnetism represents a distinct form of collinear magnetic order: it is globally compensated in real space (yielding zero net magnetization) while nevertheless breaking spin-rotation symmetry and producing a momentum-dependent spin polarization and spin splitting \cite{Jungwirth2022PRX_AM, Hayami2019JPSJ_AM, Jungwirth2022PRX_AMperspective,Igor2022PRX_AMeditorial,Yao2024AFM_AM,Han2025NRM_AM,Cheng2025AM_AM,Smejkal2025Newton_AM,Liu2025arXiv_AMReview}. An ultracold-atom realization would offer a complementary route to probe altermagnetic signatures in a highly controllable environment and to connect them directly to measurable correlation functions.

In solid-state materials, altermagnetism has so far been discussed primarily in an electronic (fermionic) context, where many transport signatures arise from the relevant spin-splitting band structure of itinerant electrons \cite{Smejkal2020, Shao2021, Naka2019}. In contrast, ultracold-atom experiments can access both fermionic and bosonic quantum matter, and it is therefore natural to ask how the altermagnetic concept manifests itself in Bose-condensed systems. This direction parallels earlier cold-atom advances that transplanted ideas initially developed for fermions to bosons, for example the implementation of synthetic spin--orbit coupling and synthetic gauge fields in Bose--Einstein condensates \cite{Lin2011Nature,Galitski2013Nature,Zhai2015RPP,Dalibard2011RMP,Goldman2014RPP}, enabling bosonic realizations of spin--momentum locking, chiral band minima, and associated collective and superfluid phenomena. Extending altermagnetism to bosons is particularly appealing because it disentangles magnetic order from charge transport and relates altermagnetic symmetry breaking directly to thermodynamic and dynamical responses. More broadly, such bosonic realizations provide a controlled setting to clarify how magnetization (local versus global) intertwines with superfluidity \cite{Liu2025AAPPS_AMSC,Liu2025PRB_AMFFLO} and to identify experimentally accessible signatures in excitation spectra and correlation functions.

A concrete route toward realizing altermagnetism with ultracold atoms in optical lattices has been proposed recently in the context of a Fermi--Hubbard model with engineered anisotropic next-nearest-neighbor tunneling, which stabilizes a $d$-wave altermagnetic phase over a broad parameter regime and leads to a characteristic anisotropic spin transport response accessible in trap-expansion measurements~\cite{Das2024PRL_AMColdAtoms}. Importantly, the underlying ingredients of this proposal---namely, the ability to tailor lattice geometries and to control hopping amplitudes and their anisotropies using established lattice-engineering techniques---do not rely on fermionic statistics. They can therefore, in principle, be transferred to bosonic gases loaded into similar optical lattices.

In the present work, we develop a continuum description of a weakly interacting altermagnetic Bose gas and analyze its elementary excitations and correlations within the Bogoliubov framework. This setting may be viewed as the long-wavelength limit of lattice-based realizations, while also providing a minimal model in which the interplay between globally compensated magnetism and superfluid order can be treated analytically. We focus on how altermagnetic symmetry breaking reshapes the low-energy collective modes and how these changes are encoded in experimentally accessible observables, such as the momentum distribution and density--spin response functions. We also include beyond-mean-field effects by evaluating the corresponding Lee--Huang--Yang correction for the ground-state energy.

\section{Theory}

We consider a two-component Bose gas in the presence of altermagnetism. Denoting the intra-species interaction strengths by $g_{\uparrow\uparrow}$ and $g_{\downarrow\downarrow}$, and the inter-species interaction by $g_{\uparrow\downarrow}=g_{\downarrow\uparrow}$, the binary Bose mixture in free space is described (in units with $\hbar=1$ and volume $\mathcal{V}=1$) by
\begin{equation}
  \hat{\mathcal{H}} = \hat{\mathcal{H}}_{0} + \hat{\mathcal{H}}_{\mathrm{int}}.
\end{equation}
The noninteracting Hamiltonian reads
\begin{equation}
  \hat{\mathcal{H}}_{0} = \sum_{\sigma=\uparrow,\downarrow} \sum_{\mathbf{k}} \epsilon_{\sigma}(\mathbf{k})\, \hat{a}_{\mathbf{k}\sigma}^{\dagger}\hat{a}_{\mathbf{k}\sigma},
\end{equation}
with the single-particle dispersion
\begin{equation}
  \epsilon_{\sigma}(\mathbf{k}) = \epsilon(\mathbf{k}) + s_{\sigma} J_{\mathbf{k}}.
\end{equation}
Here $\epsilon(\mathbf{k})=k^{2}/(2m)$, where $m$ is the atomic mass, and
\begin{equation}
  J_{\mathbf{k}} = \lambda\,\frac{k_{x}^{2}-k_{y}^{2}}{2m}
\end{equation}
describes a $d_{x^{2}-y^{2}}$-wave (altermagnetic) spin splitting. We take $s_{\uparrow}=+1$ and $s_{\downarrow}=-1$, and $\lambda$ ($0\le\lambda\le 1$) controls the strength of the altermagnetic term.

The interaction Hamiltonian is
\begin{equation}
  \hat{\mathcal{H}}_{\mathrm{int}} = \frac{1}{2}\sum_{\sigma,\sigma^{\prime}=\uparrow,\downarrow} g_{\sigma\sigma^{\prime}}
  \sum_{\mathbf{k},\mathbf{k}^{\prime},\mathbf{q}}
  \hat{a}_{\mathbf{k}+\mathbf{q},\sigma}^{\dagger}
  \hat{a}_{\mathbf{k}^{\prime}-\mathbf{q},\sigma^{\prime}}^{\dagger}
  \hat{a}_{\mathbf{k}^{\prime},\sigma^{\prime}}
  \hat{a}_{\mathbf{k},\sigma}.
\end{equation}
At zero temperature, we work in the grand-canonical ensemble with
\begin{equation}
  \hat{\mathcal{K}} = \hat{\mathcal{H}} - \sum_{\sigma} \mu_{\sigma}\hat{N}_{\sigma},
\end{equation}
where $\hat{N}_{\sigma}\equiv\sum_{\mathbf{k}}\hat{a}_{\mathbf{k}\sigma}^{\dagger}\hat{a}_{\mathbf{k}\sigma}$.

\subsection{Single-particle dispersion}
\label{subsec:single_particle_dispersion}
The altermagnetic term $J_{\mathbf{k}}=\lambda\left(k_{x}^{2}-k_{y}^{2}\right)/(2m)$ can be absorbed into an anisotropic effective-mass tensor. The single-particle dispersions therefore take the explicit form
\begin{align}
  \epsilon_{\uparrow}(\mathbf{k}) &= \frac{k_{x}^{2}}{2m_{+}} + \frac{k_{y}^{2}}{2m_{-}} + \frac{k_{z}^{2}}{2m}, \\
  \epsilon_{\downarrow}(\mathbf{k}) &= \frac{k_{x}^{2}}{2m_{-}} + \frac{k_{y}^{2}}{2m_{+}} + \frac{k_{z}^{2}}{2m},
\end{align}
with the effective masses
\begin{equation}
  m_{\pm} = \frac{m}{1\pm\lambda}.
\end{equation}
For $0\le\lambda<1$, both $m_{+}$ and $m_{-}$ are positive, and the single-particle dispersions are stable and strictly convex. In particular, $\epsilon_{\sigma}(\mathbf{k})$ is minimized at $\mathbf{k}=\mathbf{0}$ and increases monotonically away from the origin, with an anisotropic curvature set by $m_{\pm}$.

\subsection{Two-particle interactions}
\label{subsec:two_particle_interactions}

In three dimensions, the bare contact couplings $g_{\sigma\sigma'}$ require ultraviolet regularization. We express them in terms of the $s$-wave scattering lengths $a_{\sigma\sigma'}$ through the renormalization condition \cite{Salasnich2016PhysRep}
\begin{equation}
  \frac{1}{g_{\sigma\sigma'}}
  = \frac{m_{\sigma\sigma'}}{4\pi a_{\sigma\sigma'}}
  - \sum_{\mathbf{k}}\frac{1}{2\varepsilon_{\sigma\sigma'}(\mathbf{k})},
  \label{eq:g_renorm_general}
\end{equation}
where $m_{\sigma\sigma'}$ is the effective (two-body) mass entering the corresponding scattering channel and $\varepsilon_{\sigma\sigma'}(\mathbf{k})$ is the relative kinetic energy (see Appendix for details).

For the inter-species channel, we take the isotropic free-particle dispersion $\varepsilon_{\uparrow\downarrow}(\mathbf{k})\equiv\epsilon(\mathbf{k})=k^{2}/(2m)$ and set $m_{\uparrow\downarrow}=m$, which yields
\begin{equation}
  \frac{1}{g_{\uparrow\downarrow}}
  = \frac{m}{4\pi a_{\uparrow\downarrow}} - \sum_{\mathbf{k}}\frac{1}{2\epsilon(\mathbf{k})}.
  \label{eq:g_ud_renorm}
\end{equation}

For the intra-species channels, the altermagnetic single-particle dispersion implies an anisotropic quadratic kinetic term. It is convenient to introduce a single effective mass scale
\begin{equation}
  m_{0} \equiv \gamma_{0} m, \qquad \gamma_{0}=\left(1-\lambda^{2}\right)^{-1/3},
\end{equation}
which corresponds to the geometric mean of the directional masses in Subsec.~\ref{subsec:single_particle_dispersion}. We then take $m_{\uparrow\uparrow}=m_{\downarrow\downarrow}\equiv m_{0}$ and write, for $\sigma\in\{\uparrow,\downarrow\}$,
\begin{equation}
  \frac{1}{g_{\sigma\sigma}}
  = \frac{m_{0}}{4\pi a_{\sigma\sigma}} - \sum_{\mathbf{k}}\frac{1}{2\epsilon_{\sigma}(\mathbf{k})}.
  \label{eq:g_ss_renorm}
\end{equation}

For later convenience, we define the corresponding low-energy couplings,
\begin{equation}
  \tilde{g}_{\uparrow\downarrow} \equiv \frac{4\pi a_{\uparrow\downarrow}}{m},
  \qquad
  \tilde{g}_{\sigma\sigma} \equiv \frac{4\pi a_{\sigma\sigma}}{m_{0}}.
\end{equation}
In the weak-coupling regime considered in this work, Eqs.~\eqref{eq:g_renorm_general}--\eqref{eq:g_ss_renorm} imply the expansion
\begin{equation}
  g_{\sigma\sigma'} \simeq \tilde{g}_{\sigma\sigma'}
  + \tilde{g}_{\sigma\sigma'}^{2}\sum_{\mathbf{k}}\frac{1}{2\varepsilon_{\sigma\sigma'}(\mathbf{k})},
  \label{eq:g_weak_expansion}
\end{equation}
with $\varepsilon_{\uparrow\downarrow}(\mathbf{k})\equiv\epsilon(\mathbf{k})$ and $\varepsilon_{\sigma\sigma}(\mathbf{k})\equiv\epsilon_{\sigma}(\mathbf{k})$.

\subsection{Mean-field solution}
\label{subsec:mean_field_solution}

The minima of the single-particle dispersions occur at zero momentum. We therefore assume that both spin components condense at $\mathbf{k}=\mathbf{0}$,
\begin{equation}
  \langle \hat{a}_{0\sigma} \rangle \equiv \sqrt{n_{\sigma}},
\end{equation}
where $n_{\sigma}$ is the condensate density of component $\sigma$. Replacing the condensate operators by $c$-numbers and retaining terms up to quadratic order in the fluctuation operators $\hat{a}_{\mathbf{k}\sigma}$ with $\mathbf{k}\neq\mathbf{0}$ leads to the Bogoliubov (Gaussian) theory \cite{Pitaevskii2016BEC,Larsen1963AnnPhys}.

The mean-field contribution to the grand potential is
\begin{equation}
  \tilde{\mathcal{K}}^{(0)}
  = \frac{1}{2}\sum_{\sigma,\sigma'} \tilde{g}_{\sigma\sigma'}\,n_{\sigma}n_{\sigma'}
  - \sum_{\sigma} \mu_{\sigma}n_{\sigma}.
\end{equation}
Minimizing with respect to $n_{\sigma}$ yields the chemical potentials
\begin{align}
  \mu_{\uparrow} &= \tilde{g}_{\uparrow\uparrow}n_{\uparrow}+\tilde{g}_{\uparrow\downarrow}n_{\downarrow},\\
  \mu_{\downarrow} &= \tilde{g}_{\downarrow\uparrow}n_{\uparrow}+\tilde{g}_{\downarrow\downarrow}n_{\downarrow}.
\end{align}
The homogeneous mixture is mechanically stable and miscible provided that
\begin{equation}
  \tilde g_{\uparrow\uparrow}>0,\qquad \tilde g_{\downarrow\downarrow}>0,\qquad \tilde g_{\uparrow\downarrow}^{2}<\tilde g_{\uparrow\uparrow}\tilde g_{\downarrow\downarrow}.
\end{equation}
For equal intraspecies scattering lengths $a_{\uparrow\uparrow}=a_{\downarrow\downarrow}\equiv a>0$ and $a_{\uparrow\downarrow}\equiv a_{12}$, this condition can be expressed as
\begin{equation}
  \left|a_{12}\right| < (1-\lambda^2)^{1/3}a.
\end{equation}
Thus, for all positive scattering lengths, a smaller interspecies scattering length $a_{12}$ is needed to avoid phase separation in the presence of altermagnetism.

For each nonzero momentum $\mathbf{k}$, the quadratic grand-canonical Hamiltonian can be written in Nambu form,
\begin{equation}
  \hat{\mathcal{K}}^{(2)}
  = \frac{1}{2}\sum_{\mathbf{k}\neq \mathbf{0}}
  \hat{\Phi}_{\mathbf{k}}^{\dagger}\,\mathcal{H}_{\mathbf{k}}\,\hat{\Phi}_{\mathbf{k}},
\end{equation}
with the Nambu spinor
\begin{equation}
  \hat{\Phi}_{\mathbf{k}}^{\dagger}
  = \bigl(\hat{a}_{\mathbf{k}\uparrow}^{\dagger},\,\hat{a}_{\mathbf{k}\downarrow}^{\dagger},\,\hat{a}_{-\mathbf{k}\uparrow},\,\hat{a}_{-\mathbf{k}\downarrow}\bigr).
\end{equation}
The Bogoliubov matrix has the block structure
\begin{equation}
  \mathcal{H}_{\mathbf{k}}=
  \begin{pmatrix}
    A_{\mathbf{k}} & B_{\mathbf{k}}\\
    B_{\mathbf{k}} & A_{\mathbf{k}}
  \end{pmatrix},
\end{equation}
where
\begin{align}
  A_{\mathbf{k}} &=
  \begin{pmatrix}
    \epsilon_{\uparrow}(\mathbf{k})+\tilde{g}_{\uparrow\uparrow}n_{\uparrow} & \tilde{g}_{\uparrow\downarrow}\sqrt{n_{\uparrow}n_{\downarrow}}\\
    \tilde{g}_{\uparrow\downarrow}\sqrt{n_{\uparrow}n_{\downarrow}} & \epsilon_{\downarrow}(\mathbf{k})+\tilde{g}_{\downarrow\downarrow}n_{\downarrow}
  \end{pmatrix},\\
  B_{\mathbf{k}} &=
  \begin{pmatrix}
    \tilde{g}_{\uparrow\uparrow}n_{\uparrow} & \tilde{g}_{\uparrow\downarrow}\sqrt{n_{\uparrow}n_{\downarrow}}\\
    \tilde{g}_{\uparrow\downarrow}\sqrt{n_{\uparrow}n_{\downarrow}} & \tilde{g}_{\downarrow\downarrow}n_{\downarrow}
  \end{pmatrix}.
\end{align}

We introduce the Bogoliubov transformation
\begin{equation}
  \hat{a}_{\mathbf{k}\sigma}
  = \sum_{\eta=\pm}\left[u_{\sigma\eta}(\mathbf{k})\,\hat{b}_{\eta\mathbf{k}} + v_{\sigma\eta}(\mathbf{k})\,\hat{b}^{\dagger}_{\eta,-\mathbf{k}}\right],
\end{equation}
with the normalization condition
\begin{equation}
  \sum_{\sigma}\left(\left|u_{\sigma\eta}(\mathbf{k})\right|^{2}-\left|v_{\sigma\eta}(\mathbf{k})\right|^{2}\right)=1.
\end{equation}
Diagonalizing $\mathcal{H}_{\mathbf{k}}$ yields the coupled Bogoliubov--de Gennes equations
\begin{align}
  A_{\mathbf{k}}\,\vec{u}_{\eta}(\mathbf{k}) + B_{\mathbf{k}}\,\vec{v}_{\eta}(\mathbf{k}) &= \omega_{\eta}(\mathbf{k})\,\vec{u}_{\eta}(\mathbf{k}),\\
  A_{\mathbf{k}}\,\vec{v}_{\eta}(\mathbf{k}) + B_{\mathbf{k}}\,\vec{u}_{\eta}(\mathbf{k}) &= -\omega_{\eta}(\mathbf{k})\,\vec{v}_{\eta}(\mathbf{k}),
\end{align}
where $\vec{u}_{\eta}(\mathbf{k})=\bigl[u_{\uparrow\eta}(\mathbf{k}),u_{\downarrow\eta}(\mathbf{k})\bigr]^T$ and similarly for $\vec{v}_{\eta}(\mathbf{k})$.

The two Bogoliubov branches $\omega_{\pm}(\mathbf{k})$ can be written as
\begin{widetext}
  \begin{align}
    \omega_{\pm}^{2}(\mathbf{k})
    &= \frac{1}{2}\Bigl[\omega_{\uparrow}^{2}(\mathbf{k})+\omega_{\downarrow}^{2}(\mathbf{k})
      \pm \sqrt{\bigl(\omega_{\uparrow}^{2}(\mathbf{k})-\omega_{\downarrow}^{2}(\mathbf{k})\bigr)^{2}
    +16\tilde{g}_{\uparrow\downarrow}^{2}n_{\uparrow}n_{\downarrow}\,\epsilon_{\uparrow}(\mathbf{k})\epsilon_{\downarrow}(\mathbf{k})}\Bigr],
    \label{eq:omega_pm}
  \end{align}

  with
  \begin{equation}
    \omega_{\sigma}^{2}(\mathbf{k})=\epsilon_{\sigma}(\mathbf{k})\bigl[\epsilon_{\sigma}(\mathbf{k})+2\tilde{g}_{\sigma\sigma}n_{\sigma}\bigr].
    \label{eq:omega_sigma}
  \end{equation}

  \section{Results}
  \subsection{Sound velocity}
  Hereafter, unless stated otherwise, we focus on the symmetric case $\tilde{g}_{\uparrow\uparrow}=\tilde{g}_{\downarrow\downarrow}\equiv g$, $\tilde{g}_{\uparrow\downarrow}\equiv g^{\prime}$, and $n_{\uparrow}=n_{\downarrow}\equiv n/2$, where $n$ is the total density. In the long-wavelength limit $|\mathbf{k}|\to 0$, the excitations become linear, $\omega_{\pm}(\mathbf{k})\simeq c_{\pm}(\theta,\phi;\lambda)\,k$, with anisotropic sound velocities
  \begin{equation}
    c_{\pm}(\theta,\phi;\lambda)=\sqrt{\frac{n}{2m}\left(g\pm\sqrt{g^{\prime 2}+\lambda^{2}\left(g^{2}-g^{\prime 2}\right)\sin^{4}\theta\,\cos^{2}(2\phi)}\right)}.
    \label{eq:sound_speed_pm}
  \end{equation}
\end{widetext}

For $\lambda=0$, the sound velocities reduce to the usual isotropic results,
\begin{equation}
  \tilde c_{\pm}=\sqrt{\frac{n}{2m}\left(g\pm g^{\prime}\right)}.
\end{equation}
Notably, the combination $c_{+}^{2}+c_{-}^{2}=ng/m$ is isotropic and independent of $\lambda$, while each branch separately exhibits a pronounced angular dependence.

\begin{figure}[t]
  \centering
  \includegraphics[width=0.48\textwidth]{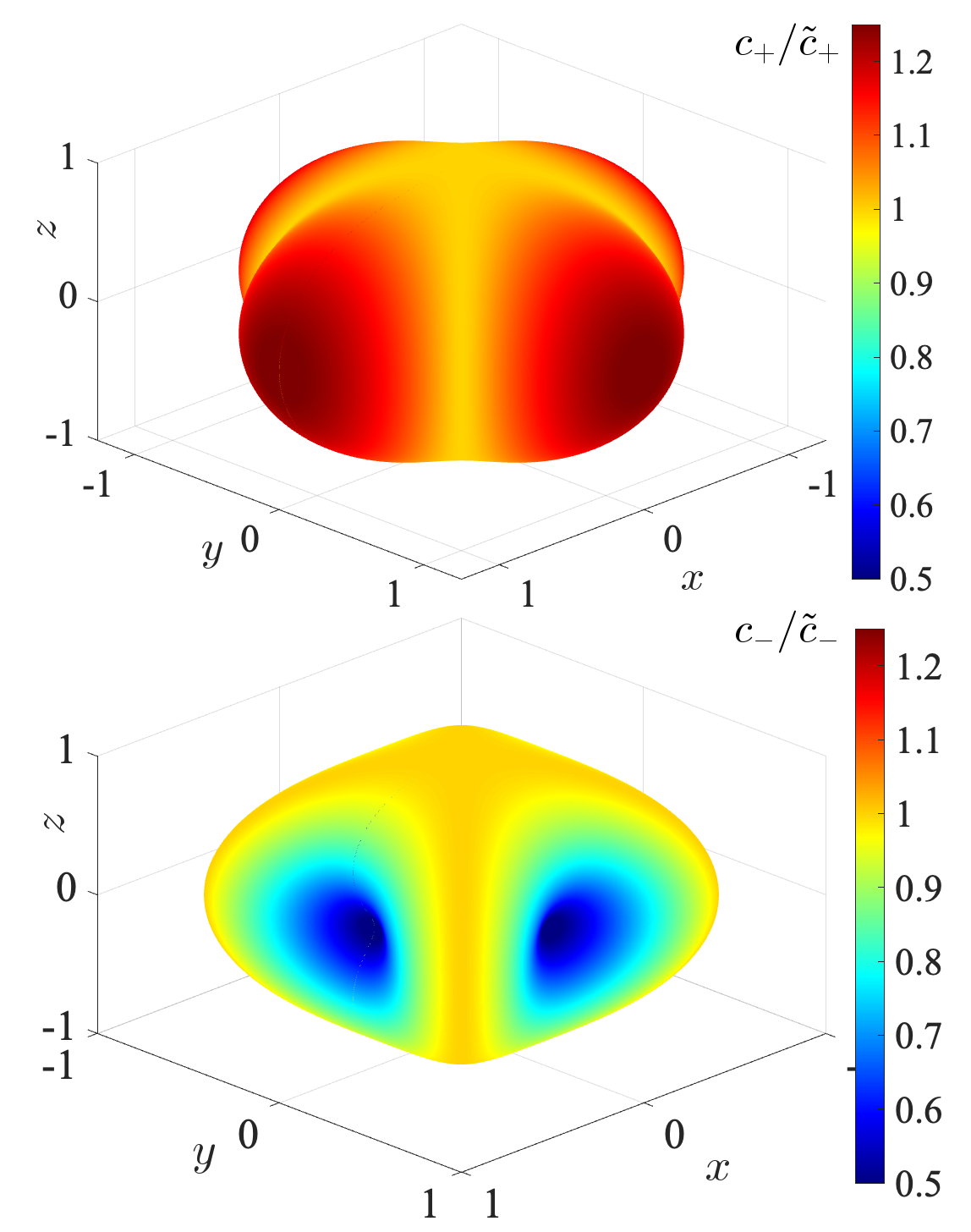}
  \caption{Angular dependence of the sound velocities in the altermagnetic Bose condensate. The color and radial distance of each point represent the sound speeds $c_{\pm}(\theta,\phi;\lambda)$ for propagation along the direction $(\theta,\phi)$, normalized by the isotropic reference values $\tilde c_{\pm}\equiv c_{\pm}(\lambda=0)$. Parameters: $na^3= 0.002$, $a_{\uparrow\downarrow}=0.1a$, and $\lambda=0.8$.}
  \label{fig:sound_velocity_anisotropy}
\end{figure}

Figure~\ref{fig:sound_velocity_anisotropy} visualizes the anisotropy induced by the $d_{x^{2}-y^{2}}$ altermagnetic splitting $J_{\mathbf{k}}\propto k_{x}^{2}-k_{y}^{2}$. The effect is strongest in the $x$--$y$ plane and vanishes along directions satisfying $\cos(2\phi)=0$.

\subsection{Quantum depletion}
\begin{figure}[t]
  \centering
  \includegraphics[width=0.48\textwidth]{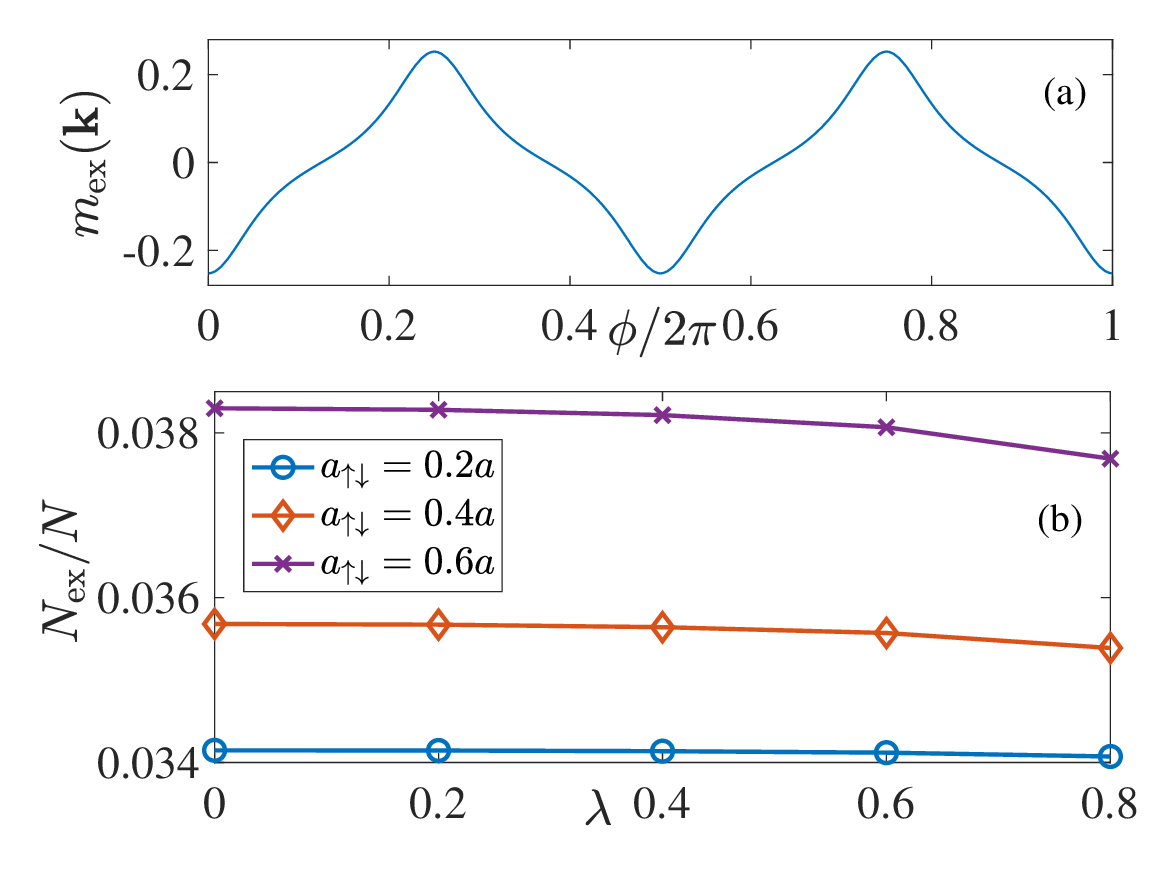}
  \caption{(a) Momentum-resolved magnetization of the quantum depletion at $k=0.1 a^{-1}$ in the $x$--$y$ plane ($\theta=\pi/2$) as a function of the azimuthal angle $\phi$. (b) Total quantum depletion as a function of $\lambda$ for different interspecies scattering lengths $a_{\uparrow\downarrow}$. The density is fixed to $na^3= 0.002$.}
  \label{fig:quantum_depletion}
\end{figure}

At the Bogoliubov level, the momentum-resolved quantum depletion of component $\sigma$ is
\begin{equation}
  n_{{\rm ex}}^{(\sigma)}(\mathbf{k})=\sum_{\eta=\pm}\left|v_{\sigma\eta}(\mathbf{k})\right|^{2}.
\end{equation}
We further define the total depletion density $n_{\rm ex}(\mathbf{k})\equiv n_{{\rm ex}}^{(\uparrow)}(\mathbf{k})+n_{{\rm ex}}^{(\downarrow)}(\mathbf{k})$ and the depletion magnetization
\begin{equation}
  m_{\rm ex}(\mathbf{k})\equiv n_{{\rm ex}}^{(\uparrow)}(\mathbf{k})-n_{{\rm ex}}^{(\downarrow)}(\mathbf{k}).
\end{equation}
Figure~\ref{fig:quantum_depletion}(a) shows that $m_{\rm ex}(\mathbf{k})$ is generally nonzero and inherits the angular structure imposed by altermagnetism. However, upon integrating over the azimuthal angle $\phi$ the magnetization averages to zero, consistent with globally compensated (net-zero) magnetization.

The total quantum depletion is given by $N_{\rm ex}=\sum_{\mathbf{k}} n_{\rm ex}(\mathbf{k})$ (replaced by $\int d\mathbf{k}/(2\pi)^{3}$ in the numerical evaluation). As shown in Fig.~\ref{fig:quantum_depletion}(b), the total depletion depends only weakly on $\lambda$ over the parameter range considered.

\subsection{Structure factor}
\label{subsec:structure_factor}

In addition to the excitation spectrum and depletion, the altermagnetic order can be diagnosed through density and spin correlations. We consider the Fourier components of the density operator,
\begin{equation}
  \hat{n}_{\sigma}(\mathbf{q})\equiv\sum_{\mathbf{k}} \hat{a}^{\dagger}_{\mathbf{k}-\mathbf{q},\sigma}\hat{a}_{\mathbf{k}\sigma}, \qquad \mathbf{q}\neq\mathbf{0}.
\end{equation}
Keeping only the leading (linear) fluctuation terms in the condensate gives
\begin{equation}
  \hat{n}_{\sigma}(\mathbf{q}) \simeq \sqrt{n_{\sigma}}\left(\hat{a}_{\mathbf{q}\sigma}+\hat{a}^{\dagger}_{-\mathbf{q}\sigma}\right).
\end{equation}
Using the Bogoliubov transformation, this can be rewritten as
\begin{equation}
  \hat{n}_{\sigma}(\mathbf{q}) \simeq \sum_{\eta=\pm} x_{\sigma\eta}(\mathbf{q})\left(\hat{b}_{\eta\mathbf{q}}+\hat{b}^{\dagger}_{\eta,-\mathbf{q}}\right),
\end{equation}
where $x_{\sigma\eta}(\mathbf{q}) \equiv \sqrt{n_{\sigma}}\left[u_{\sigma\eta}(\mathbf{q})+v_{\sigma\eta}(\mathbf{q})\right]$.

We then introduce the total-density and spin-density operators
\begin{equation}
  \hat{d}(\mathbf{q})\equiv \hat{n}_{\uparrow}(\mathbf{q})+\hat{n}_{\downarrow}(\mathbf{q}),
  \qquad
  \hat{s}(\mathbf{q})\equiv \hat{n}_{\uparrow}(\mathbf{q})-\hat{n}_{\downarrow}(\mathbf{q}).
\end{equation}
The dynamical structure factors are defined as \cite{Pitaevskii2016BEC}
\begin{equation}
  S_{AB}(\mathbf{q},\omega)\equiv \frac{1}{2\pi n}\int_{-\infty}^{\infty}dt\,e^{i\omega t}\,\langle \hat{A}(\mathbf{q},t)\hat{B}(-\mathbf{q},0)\rangle,
\end{equation}
where $n$ is the total density and $A,B\in\{d,s\}$.

\begin{figure}[t]
  \centering
  \includegraphics[width=0.48\textwidth]{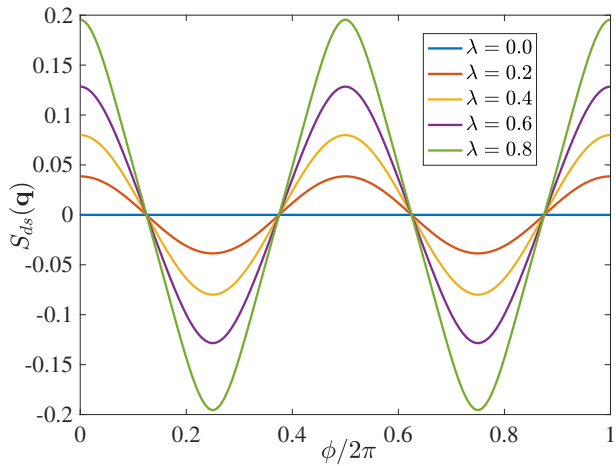}
  \caption{Mixed density--spin structure factor in the $x$--$y$ plane. Angular dependence of the static mixed structure factor $S_{sd}(\mathbf{q})$ at fixed $q=0.1a^{-1}$ and $\theta=\pi/2$ as a function of the azimuthal angle $\phi$ for various $\lambda$.}
  \label{fig:structure_factor}
\end{figure}

For $n_{\uparrow}=n_{\downarrow}$, we obtain
\begin{align}
  S_{dd}(\mathbf{q},\omega) &= \frac{1}{2}\sum_{\eta=\pm}\left[x_{\uparrow\eta}(\mathbf{q})+x_{\downarrow\eta}(\mathbf{q})\right]^{2}\,\delta\!\left[\omega-\omega_{\eta}(\mathbf{q})\right],\\
  S_{ss}(\mathbf{q},\omega) &= \frac{1}{2}\sum_{\eta=\pm}\left[x_{\uparrow\eta}(\mathbf{q})-x_{\downarrow\eta}(\mathbf{q})\right]^{2}\,\delta\!\left[\omega-\omega_{\eta}(\mathbf{q})\right],\\
  S_{sd}(\mathbf{q},\omega) &= \frac{1}{2}\sum_{\eta=\pm}\left[x_{\uparrow\eta}^{2}(\mathbf{q})-x_{\downarrow\eta}^{2}(\mathbf{q})\right]
  \,\delta\!\left[\omega-\omega_{\eta}(\mathbf{q})\right].
\end{align}
The corresponding static structure factors follow by integrating over frequency,
\begin{align}
  S_{dd}(\mathbf{q}) &= \frac{1}{2}\sum_{\eta=\pm}\left[x_{\uparrow\eta}(\mathbf{q})+x_{\downarrow\eta}(\mathbf{q})\right]^{2},\\
  S_{ss}(\mathbf{q}) &= \frac{1}{2}\sum_{\eta=\pm}\left[x_{\uparrow\eta}(\mathbf{q})-x_{\downarrow\eta}(\mathbf{q})\right]^{2},\\
  S_{sd}(\mathbf{q}) &= \frac{1}{2}\sum_{\eta=\pm}\left[x_{\uparrow\eta}^{2}(\mathbf{q})-x_{\downarrow\eta}^{2}(\mathbf{q})\right].
\end{align}

The mixed structure factor $S_{sd}(\mathbf{q})$ is identically zero in the absence of altermagnetism but becomes finite for $\lambda>0$, reflecting a local mixing between density and spin fluctuations. Figure~\ref{fig:structure_factor} shows that $S_{sd}(\mathbf{q})$ is strongly anisotropic in $\phi$; however, its angular average vanishes, consistent with globally vanished magnetization.

\subsection{Lee-Huang-Yang correction}

\begin{figure}[t]
  \centering
  \includegraphics[width=0.48\textwidth]{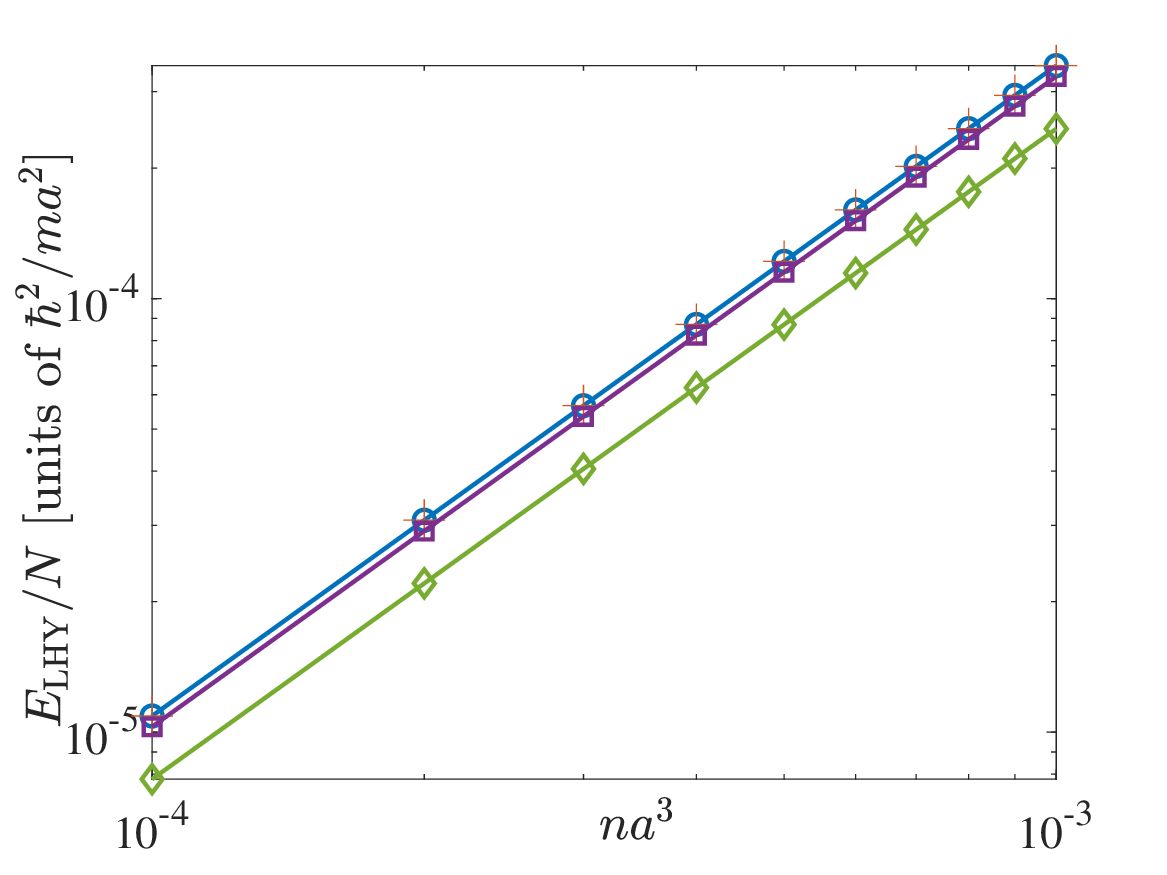}
  \caption{Lee--Huang--Yang (LHY) energy correction per particle as a function of density for several altermagnetic strengths $\lambda$. Circle, square, and diamond symbols show our numerical evaluation for $\lambda=0.0$, $0.4$, and $0.8$, respectively; crosses indicate the known analytic result in the non-altermagnetic limit $\lambda=0$. Parameters: $a_{\uparrow\downarrow}=0.1a$.}
  \label{fig:LHY_correction}
\end{figure}

Finally, we calculate the beyond mean-field LHY correctiosn of the BEC energy, which is given by
\begin{widetext}
  \begin{equation}
    E_{{\rm LHY}}=\int\frac{d^{3}k}{2\left(2\pi\right)^{3}}\left[\omega_{+}(\mathbf k)+\omega_{-} (\mathbf k)-\epsilon_{\uparrow}(\mathbf k)-\epsilon_{\downarrow}(\mathbf k)-\tilde g_{\uparrow \uparrow} n_\uparrow - \tilde g_{\downarrow \downarrow} n_\downarrow+\frac{\tilde g_{\uparrow\uparrow}^{2}n_\uparrow^{2}}{2\epsilon_{\uparrow}(\mathbf k)}+\frac{\tilde g_{\downarrow\downarrow}^{2}n_\downarrow^{2}}{2\epsilon_{\downarrow}(\mathbf k)}+\frac{\tilde g_{\uparrow\downarrow}^{2}n_\uparrow n_\downarrow}{\epsilon_{\uparrow\downarrow}(\mathbf k)}\right].
  \end{equation}
\end{widetext}
We have tested that the integrant does not have ultraviolet divergence, and the results are shown in Fig. \ref{fig:LHY_correction}. The numerical results for $\lambda = 0$ agrees with the analytical expression for non-altermagnetic case \cite{Petrov2016PRL_QD,Liu2020PRL_QDpairing,Liu2020PRA_QDpairing}. For non-zero $\lambda$, we can see that the LHY correction has a similar density dependency $E_{\rm LHY}\sim n^{5/2}$. This implies the quantum droplet can still be stablized by the LHY correction in the unstable regime, in the presence of altermagnetism.
\section{Summary}

We have developed a Bogoliubov theory for a weakly interacting two-component Bose--Einstein condensate in the presence of altermagnetism, modeled by a $d_{x^{2}-y^{2}}$-wave spin splitting in momentum space. In the stable and miscible regime, we find that altermagnetism imprints a pronounced angular dependence on low-energy collective modes: the two sound velocities become anisotropic, while their squared sum remains isotropic and independent of the altermagnetic coupling strength.

Beyond the excitation spectrum, we have analyzed correlation functions that are directly accessible in ultracold-atom experiments. The momentum-resolved quantum depletion exhibits a nontrivial local magnetization pattern that averages to zero upon angular integration, consistent with globally compensated magnetism. Similarly, the mixed density--spin structure factor becomes finite and anisotropic for nonzero altermagnetism, yet its angular average vanishes.

Finally, we have evaluated the Lee--Huang--Yang correction in the altermagnetic phase and found that it retains the characteristic $n^{5/2}$ scaling with density, indicating that beyond-mean-field quantum fluctuations remain important and can, in principle, stabilize self-bound droplets in appropriate interaction regimes \cite{Petrov2016PRL_QD,Liu2020PRL_QDpairing,Liu2020PRA_QDpairing}. The droplet formation with altermagnetsim will be addressed in future studies.

Our results provide a minimal theoretical framework for altermagnetic Bose superfluids that could be realized in optical lattices at low filling. In the weakly interacting regime, the theory gives quantitative predictions for the anisotropic angular dependence of low-energy excitations induced by altermagnetism. Although such measurements can be challenging because they require spin- and momentum-resolved detection, the predicted anisotropic sound velocities may be probed by launching a moving density dip using a dimple optical potential along different directions, following the technique demonstrated with sodium-23 atoms \cite{Andrews1997PRL_SoundBEC}.

\begin{acknowledgments}
  This research was supported by the Australian Research Council's (ARC) Discovery Program, Grants Nos. FT230100229 (J.W.), DP240100248 (X.-J.L.), and DP240101590 (H.H.).
\end{acknowledgments}

\appendix
\section{Regularization of contact interactions}
\label{sec:appendix_regularization}

In this Appendix we briefly describe the ultraviolet regularization used for the three-dimensional contact couplings in Sec.~\ref{subsec:two_particle_interactions}. It is convenient to work in real space, where the single-particle Hamiltonians read
\begin{align}
  \hat{h}_{\uparrow} &= -\frac{1}{2m_{+}}\frac{\partial^{2}}{\partial x^{2}}-\frac{1}{2m_{-}}\frac{\partial^{2}}{\partial y^{2}}-\frac{1}{2m}\frac{\partial^{2}}{\partial z^{2}},\\
  \hat{h}_{\downarrow} &= -\frac{1}{2m_{-}}\frac{\partial^{2}}{\partial x^{2}}-\frac{1}{2m_{+}}\frac{\partial^{2}}{\partial y^{2}}-\frac{1}{2m}\frac{\partial^{2}}{\partial z^{2}}.
\end{align}

For an $\uparrow$ atom and a $\downarrow$ atom, introducing the relative coordinate $\mathbf{x}\equiv\mathbf{x}_{2}-\mathbf{x}_{1}$ leads to the relative Hamiltonian
\begin{equation}
  \hat{h}_{\rm rel}^{(\uparrow\downarrow)}=-\frac{1}{2\mu_{\uparrow\downarrow}}\nabla^{2}+g_{\uparrow\downarrow}\,\delta(\mathbf{x}),
\end{equation}
with reduced mass $\mu_{\uparrow\downarrow}=m/2$. Owing to the symmetry between the two spin components, the relative kinetic energy is isotropic and the regularization coincides with the standard non-altermagnetic result, Eq.~\eqref{eq:g_ud_renorm}.

For two identical spins the relative kinetic term is anisotropic. For example, for two $\uparrow$ atoms one finds
\begin{equation}
  \hat{h}_{\rm rel}^{(\uparrow\uparrow)}=-\frac{1}{m_{+}}\frac{\partial^{2}}{\partial x^{2}}-\frac{1}{m_{-}}\frac{\partial^{2}}{\partial y^{2}}-\frac{1}{m}\frac{\partial^{2}}{\partial z^{2}}+g_{\uparrow\uparrow}\,\delta(\mathbf{x}).
\end{equation}
It is convenient to rescale coordinates,
\begin{equation}
  \tilde x = \sqrt{\frac{m_{+}}{m_{0}}}\,x,\qquad
  \tilde y = \sqrt{\frac{m_{-}}{m_{0}}}\,y,\qquad
  \tilde z = \sqrt{\frac{m}{m_{0}}}\,z,
\end{equation}
where $m_{0}\equiv(m_{+}m_{-}m)^{1/3}=m\left(1-\lambda^{2}\right)^{-1/3}$ is the geometric-mean mass used in Sec.~\ref{subsec:two_particle_interactions}. In these variables the kinetic energy becomes isotropic, and one may write schematically
\begin{equation}
  \tilde{h}_{\rm rel}^{(\uparrow\uparrow)}=-\frac{1}{m_{0}}\tilde{\nabla}^{2}+g_{\uparrow\uparrow}\,\delta(\tilde{\mathbf{x}}),
\end{equation}
which yields the standard three-dimensional renormalization condition
\begin{equation}
  \frac{1}{g_{\uparrow\uparrow}}
  = \frac{m_{0}}{4\pi a_{\uparrow\uparrow}} - \sum_{\tilde{\mathbf{k}}}\frac{1}{2\epsilon_{0}(\tilde{\mathbf{k}})}.
  \label{eq:g_uu_renorm}
\end{equation}
Here $\epsilon_{0}(\tilde{\mathbf{k}})\equiv \tilde{k}^{2}/(2m_{0})$. Re-expressing the momentum sum in the original variables reproduces Eq.~\eqref{eq:g_ss_renorm} in the main text (and similarly for the $\downarrow\downarrow$ channel).

\bibliography{AMBEC}
\end{document}